\documentclass[journal=apchd5,manuscript=article]{achemso}

\usepackage{amsmath}
\usepackage{amssymb}

\author{Artsiom Kazlou}
\affiliation{Faculty of Physics, University of Bialystok, 15-245 Bialystok, Poland}
\author{Alexander L. Chekhov}
\affiliation{Department of Physics, Freie Universit{\"a}t Berlin, 14195 Berlin, Germany}
\author{Alexander I. Stognij}
\affiliation{Scientific-Practical Materials Research Centre of the NASB, 220072 Minsk, Belarus}
\author{Ilya Razdolski}
\author{Andrzej Stupakiewicz}
\email{and@uwb.edu.pl}
\affiliation{Faculty of Physics, University of Bialystok, 15-245 Bialystok, Poland}

\title[]
  {Surface plasmon-enhanced photo-magnetic excitation of spin dynamics in Au/YIG:Co magneto-plasmonic crystals}

\begin{document}

\begin{abstract}
We report strong amplification of photo-magnetic spin precession in Co-doped YIG employing a surface plasmon excitation in a metal-dielectric magneto-plasmonic crystal. Plasmonic enhancement is accompanied by the localization of the excitation within the 300~nm-thick layer inside the transparent dielectric garnet. Experimental results are nicely reproduced by numerical simulations of the photo-magnetic excitation. Our findings demonstrate the magneto-plasmonic concept of subwavelength localization and amplification of the photo-magnetic excitation in dielectric YIG:Co and open up a path to all-optical magnetization switching below diffraction limit with energy efficiency approaching the fundamental limit for magnetic memories. 
\end{abstract}

In the last decade, the rapid progress of ultrafast optomagnetism has opened up rich possibilities for optical data recording in magnetic materials, aiming at the heat-assisted magnetic recording with $20\times20\times10$ nm bit size in metallic media \cite{Challener2009}. Meanwhile, the research has been fueled into alternative approaches, including the highly promising all-optical magnetization switching demonstrated in multiple metallic systems in the last few years \cite{Stanciu2007, Fullerton}. Yet, the all-optical switching in metals requires heating to high temperatures and demagnetization \cite{KirilyukRMP}. Recently, based on a direct modification of the magnetic anisotropic energy barrier \cite{Stupakiewicz2001,Hansteen2005,Atoneche2010}, a non-thermal method for ultrafast photo-magnetic recording in dielectric garnets has been developed \cite{Stupakiewicz2017} employing magnetization precession mechanism \cite{Stupakiewicz2019, Szerenos2019}. There, the key to future applications is the optics of photo-magnetic recording with the light localization into a $\sim20$~nm-size spot, thus approaching the Landauer limit ($\sim0.25$ aJ) \cite{Landauer}. 

This challenge of confining the photoexcitation within subwavelength volumes can be addressed with magneto-plasmonics, a rapidly developing branch of modern photonics \cite{Armelles2013, Belotelov2019,Dmitriev2020}. The high potential of magneto-plasmonics for local manipulation of magnetic order with photons is already established \cite{Liu2015,Schmising2015}. On the other hand, a highly interesting class of systems emerged recently, where magnetic dielectrics are covered with gratings made of plasmonic metals (such as Au) \cite{Sepulveda2006, Wurtz2008, Belotelov2011}. The grating allows for the free-space excitation of surface plasmon polaritons (SPPs) at both interfaces of the metal \cite{Pohl2013, Razdolski2015}, whereas the transparent dielectric layer ensures low losses of the SPP excitations, in contrast to magneto-plasmonic systems with transition metal ferromagnets \cite{Temnov2010}. Tailoring the electric field distribution inside the dielectric through the metal-bound SPP excitation enables novel nonlinear-optical and opto-magnetic effects \cite{Khurgin2006, Krutyanskiy2015, Razdolski2016, Im2017, Ho2018, Chekhov2018, Im2019,Pae2020}.

In this work, we employ this magneto-plasmonic grating approach for amplifying the photo-magnetic spin precession in dielectric Co-doped yittrium iron garnet (YIG:Co). We observe a strong increase of the magnetization precession amplitude in the vicinity of the SPP resonance in the near-infrared. Numerical simulations of the electric field distribution show that due to the SPP-induced light localization at the interface, the specific efficiency of the excitation of the magnetization precession is enhanced 6-fold within the 300~nm active layer, as compared to the bare garnet film. Because photo-magnetic switching is a threshold effect, our results represent an important step towards nanoscale photo-magnetic data writing with femtosecond laser pulses. They highlight the rich potential of magneto-plasmonic approach for scaling down towards nm-sized magnetic bits and further improving the energy efficiency of the all-optical magnetic recording. 

Experimental studies of the SPP-induced photo-magnetic anisotropy were performed on Au/YIG:Co magneto-plasmonic crystals consisting of a 7.5~$\mu$m-thick garnet film covered with Au gratings \cite{Razdolski2019}. YIG:Co is a weakly opaque (in the near infrared spectral range) ferrimagnet with a saturation magnetization of $4\pi M_s = 80$~Gs and a Neel temperature of 455~K. The Co dopants display strong single ion anisotropy which depends on the ion’s valence state. Therefore, resonant pumping of Co electronic transitions with laser pulses enables direct access to the magnetic anisotropy and thus, the magnetization. This results in the exceptionally strong photo-magnetic effect in YIG:Co \cite{Atoneche2010}, ultimately allowing ultrafast magnetization switching with a single femtosecond laser pulse \cite{Stupakiewicz2017}. The Co doping also enhances magnetocrystalline anisotropy and the Gilbert damping $\alpha = 0.2$ \cite{Stupakiewicz2017}. The YIG:Co garnet film with a composition of Y$_{2}$CaFe$_{3.9}$Co$_{0.1}$GeO$_{12}$ was grown on a Gd$_{3}$Ga$_{5}$O$_{12}$~(001) substrate. The surface of the garnet thin film was treated with a low energy oxygen ion beam \cite{PashkevichTSF}. A 50~nm-thick Au grating with a 800~nm period (gap width 100~nm) was deposited on the garnet surface by ion-beam sputtering and perforated using FIB \cite{StognijChekhov}.

\begin{figure}[t]
  \includegraphics[width=8cm]{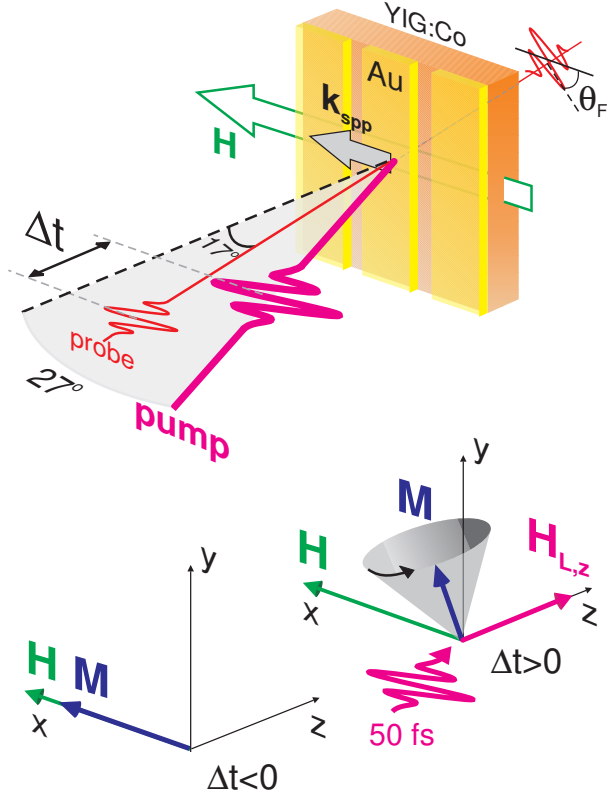}
  \caption{Schematics of the experimental pump-probe transmission geometry. The 50~fs-long near-IR {\it p}-polarized pump pulses excite a SPP resonance at the Au/YIG:Co interface. The electromagnetic SPP field induces the photo-magnetic anisotropy $\boldsymbol{\rm H_L}$ in YIG:Co triggering the magnetization precession. The latter is monitored through transient Faraday rotation $\theta_F$ of the delayed probe pulses.}
  \label{fgr:1}
\end{figure}

SPP-driven photo-magnetic excitation of magnetization precession was studied in the two-colour pump-probe transmission geometry schematically shown in Fig.~\ref{fgr:1}. There, the time-resolved Faraday rotation angle $\theta_F$ of the probe beam was monitored as a function of the delay time 
$\Delta t$ between the pump and probe pulses. The pump (probe) laser pulses with a duration of 50~fs from a Ti-Sapphire amplifier at a 500~Hz (1~kHz) repetition rate impinged at an angle of $27^{\circ}$ ($17^{\circ}$) from the sample normal, respectively, see Fig.~\ref{fgr:1}. Employing the optical parametric amplifier, the wavelength of the pump beam $\lambda$ was tuned in the near infrared range within 1200-1350~nm, while the probe wavelength was set to 800~nm. 
The more powerful pump beam was focused into a spot of about $100~\mu$m in diameter on the samples, resulting in an energy density of $\sim 4$~mJ/cm$^2$, while the spot size of the 30~times weaker probe beam was twice smaller.
Both pump and probe beams were polarized in their plane of incidence. The delay time $\Delta t$ between the pump and probe pulses was controlled by means of a motorized delay stage. 

An external magnetic field $H = 3.2$~kOe was applied in-plane of the sample along the [100] direction of the garnet crystal (Fig.~\ref{fgr:1}) to set the magnetization $\boldsymbol{\rm M}\parallel\boldsymbol{\rm H}$. This enables monitoring the magnetization precession through the transient Faraday rotation $\theta_F$ proportional to the out-of-plane component $M_z$. In this geometry, a torque $\mathcal{T}$ is exerted on $\boldsymbol{\rm M}$ by the effective field of the photo-induced magnetic anisotropy $\boldsymbol{\rm H_L}=\widehat{\chi}\;\vdots\; \boldsymbol{\rm EEM}$, where $\boldsymbol{\rm E}$ is the electric field of light and $\widehat{\chi}$ is the photo-magnetic 3rd order susceptibility tensor\cite{Stupakiewicz2017}. This torque $\mathcal{T}\propto[\boldsymbol{\rm H_L}\times \boldsymbol{\rm M}]$ triggers the magnetization precession around its new equilibrium determined by the magneto-crystalline anisotropy field $\boldsymbol{\rm H_c}$, $\boldsymbol{\rm H}$ and $\boldsymbol{\rm H_L}$. After the relaxation of $\boldsymbol{\rm H_L}$ (on the scale of 20~ps \cite{Atoneche2010, Stupakiewicz2017}), the equilibrium position for the magnetization is restored and its precession proceeds around the original effective field direction. Because the incident light pulse is {\it p}-polarized with the electric field $\boldsymbol{\rm E}=(E_x,0,E_z)$ and the garnet has the cubic $4mm$ symmetry, a non-zero torque on $M_x$  is generated by the following component of $\boldsymbol{\rm H_L}$:

\begin{equation}
    H_{L,z}=\chi_{zxzx} E_x E_z^* M_x + c.c.
    \label{eq1}
\end{equation}
Notably, because the refractive index of garnets is relatively large ($n\gtrsim 2$ in the near-infrared \cite{Landolt}), the normal-to-surface projection of the electric field $E_z$ of the propagating light is suppressed. On the contrary, being one of the characteristic features of the SPP excitation at a metal-dielectric interface, prominent enhancement of $E_z$ in the dielectric promises an amplification of the photo-magnetic anisotropy field and thus large angles of magnetization precession. 

We studied the magnetization precession in the spectral vicinity of the SPP resonance at the Au/YIG:Co interface ($\sim 1275$~nm at $27^{\circ}$ of incidence \cite{Chekhov2018,Razdolski2019}). At each pump wavelength $\lambda$, we measured the transient rotation of the probe polarization $\theta$ at the two opposite directions of $\boldsymbol{\rm H}$ and analyzed their difference, thus removing concomitant signal variations of a non-magnetic origin. To verify the symmetry and the magnitude of the photo-magnetic effect and enable a reference point, similar measurements were performed on a bare YIG:Co film without an Au grating.

\begin{figure}[t]
  \includegraphics[width=\columnwidth]{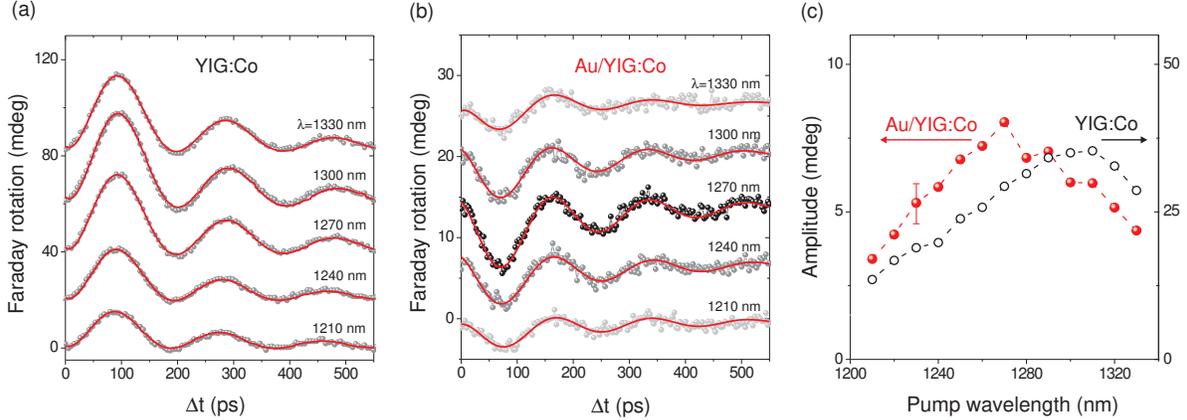}
  \caption{Time-resolved Faraday rotation in bare YIG:Co film (a) and Au/YIG:Co magneto-plasmonic crystal (b) induced by pump pulses of varied wavelength. The datasets are shifted vertically without rescaling. The red lines show the single-frequency damped sine function fitted to the data. c) Spectral dependence of the Faraday rotation amplitude extracted from the fits shown in panels (a,b).}
  \label{fgr:2}
\end{figure}

The time-resolved Faraday rotation traces  $\theta_F = [\theta (+H) - \theta (-H)]/2$ obtained on the bare garnet and the Au/YIG:Co magneto-plasmonic crystal are shown in Fig.~\ref{fgr:2}(a,b), respectively. It is seen that both samples demonstrate similar magnetization dynamics which can be reasonably well described by a single-mode precession at about 5~GHz frequency, in agreement with the previous findings \cite{Atoneche2010}. Other temporal details of the magnetization dynamics will be discussed in a subsequent publication, and here we only focus on the precession amplitude. It can be further seen in Fig.~\ref{fgr:2} that the absolute amplitude in the YIG:Co sample is approximately one order of magnitude stronger than that in the magneto-plasmonic Au:YIG:Co crystal. However, and this will be the central point of the following discussion, the spectral behavior of these two samples differs noticeably (Fig.~\ref{fgr:2}c). Indeed, in the bare garnet the precession is slightly enhanced at around $\lambda\approx1300$~nm, whereas the Au/YIG:Co exhibits a different spectral shape, with the largest precession amplitude observed around $\lambda\approx1270$~nm (accentuated with darker points in Fig.~\ref{fgr:2}b).

\begin{figure}[!h]
  \includegraphics[width=\columnwidth]{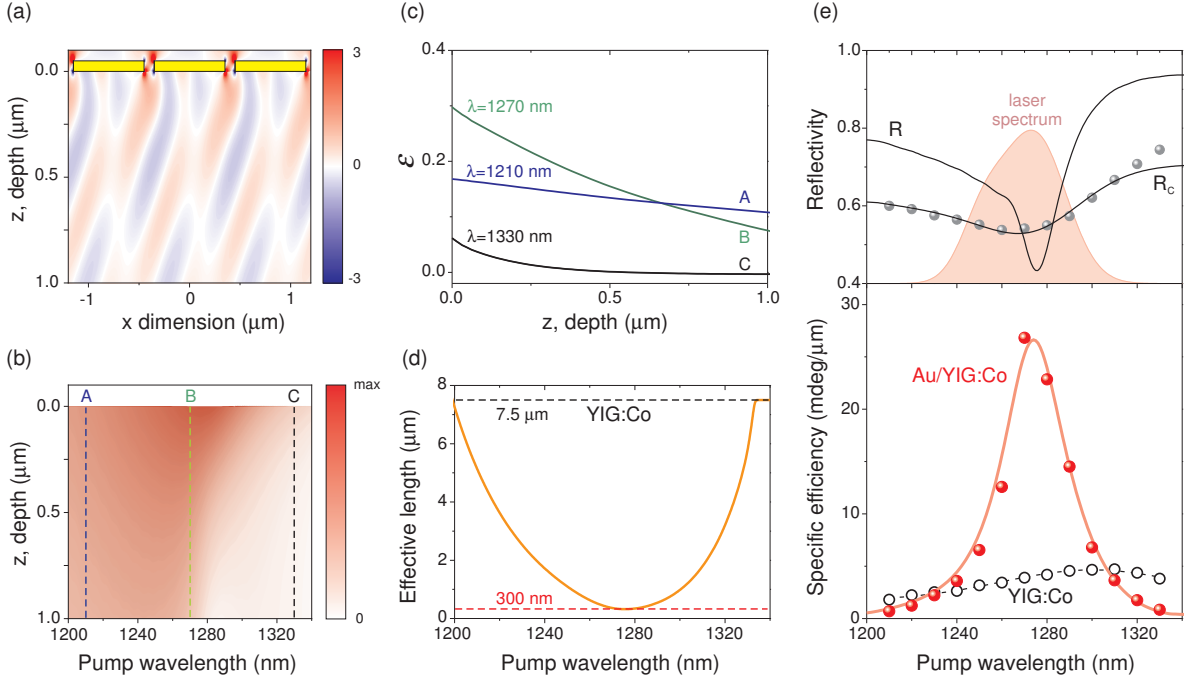}
  \caption{
  a) Numerically simulated spatial distribution $\mathcal{E}(x,z)$ inside the YIG:Co layer 
  at the SPP resonance ($\lambda=1270$~nm). The yellow bars indicate the Au grating. 
  b-c) In-depth profiles $\mathcal{E}(z,\lambda)$ 
  integrated across the $x$-dimension. The dashed lines indicate three $\lambda$ values which are shown in (c) in detail;
  $B$ ($\lambda=1270$~nm) corresponds to the resonant SPP excitation.
  d) Calculated effective decay depth $l_{\rm eff}$ of the photo-magnetic excitation away from the Au/YIG:Co interface. The top dashed line at $7.5~\mu$m indicates the total thickness of the YIG:Co layer. The bottom dashed line at $300$~nm shows the shortest effective depth obtained at the resonance.
  e) (top) Calculated reflectivity spectra, before ($R_c$) and after ($R$) taking into account the spectral broadening due to the finite laser pulse duration
  . Full dots: experimental Au/YIG:Co reflectivity. The shaded area illustrates the measured laser spectrum at around $\lambda=1270$~nm. (bottom) Specific efficiency of the photo-magnetic excitation for the bare garnet (open dots) and Au:YIG/Co (full dots). The solid red line is the result of the numerical simulations.}
  \label{fgr:3}
\end{figure}

To quantify the SPP-induced enhancement of the precession amplitude, we performed numerical simulations of the SPP-driven excitation employing dedicated Lumerical software \cite{Lumerical}. From Eq.~\ref{eq1} it can be shown that the photo-induced anisotropy field $H_L$ takes the form:
\begin{equation}
    H_L\propto \mathcal{E} \equiv E_x E_z^*+E_z E_x^*=2|E_x||E_z|\cos\varphi,
    \label{eq2}
\end{equation}
where $\varphi$ is the phase shift between the two components of the electric field $E_x$,$E_z$. We calculated $\mathcal{E}$ inside the YIG:Co layer in the vicinity of the SPP resonance. Optical constants of Au were taken from Ref.~\citenum{JohnsonChristy}. Fig.~\ref{fgr:3} shows the characteristic spatial distribution of $\mathcal{E}(x,z,\lambda)$ calculated for the $\lambda=1270$~nm, i.e. at the SPP resonance. In the Figure, we only show the top 1~$\mu$m-thick layer of YIG:Co although the calculations were performed for the entire garnet film.  These data maps were collected for a set of $\lambda$ and for each of them, integrated across the {\it x}-axes to enable the spectral comparison of the in-depth distributions $\mathcal{E}(z,\lambda)$. Those results are shown in Fig.~\ref{fgr:3}b, where the darker shaded regions illustrate the enhancement and interfacial localization of the optical excitation. To highlight the latter point, we show a few selected in-depth profiles (indicated with dashed lines in Fig.~\ref{fgr:3}b) in Fig.~\ref{fgr:3}c. At each $\lambda$, from these profiles we calculated the effective excitation depth $l_{\rm eff} =|\mathcal{E}\cdot(d\mathcal{E}/dz)^{-1}|\,_{z=0}$ containing the most significant part of the excitation energy (see in Fig.~\ref{fgr:3}d). It is seen there that at the SPP resonance, the excitation is concentrated within the 300~nm layer adjacent to the Au-garnet interface. On the contrary, away from the resonance, the effective depth $l_{\rm eff}$ increases rapidly towards the total thickness of the YIG:Co layer ($7.5~\mu$m).

Finally, in order to compare the results of our calculations with the experimental data, spectral broadening due to the finite laser pulse duration has to be taken into account. Indeed, with 50~fs-short laser pulses, every experimental wavelength shown in Fig. 2 in fact represents a continuum of wavelengths around the indicated value. A characteristic 50~nm-wide spectrum of a laser pulse $S(\lambda)$ centered around $\lambda=1270$~nm is exemplified in the top panel of Fig.~\ref{fgr:3}e with the shaded area. To verify the broadening, we compared the experimental reflectivity spectrum of the Au/YIG:Co magneto-plasmonic crystal and the calculated data $R(\lambda)$ convoluted with $S(\lambda)$: $R_c(\lambda)\equiv R(\lambda)\ast S(\lambda) = \int R(\bar{\lambda})S(\bar{\lambda}-\lambda) d\bar{\lambda}$. A very good agreement with the experimental data allows us to apply the same procedure to the calculated in-depth integrated $\mathcal{E}(\lambda)=\int \mathcal{E}(z,\lambda)dz$ data to obtain specific excitation efficiency $\xi(\lambda)=\left[\mathcal{E}(\lambda)\chi(\lambda)\right]\ast S(\lambda)/l_{\rm eff}(\lambda)$.
Here, $\chi(\lambda)$ accounts for the dispersion of the photo-magnetic tensor $\widehat{\chi}$ discussed earlier, which can be extracted from the spectral dependence of the precession amplitude obtained on a bare YIG:Co. 

The results of this procedure are summarized in the bottom panel of Fig.~\ref{fgr:3}e. There, we also show the specific excitation efficiency for the bare transparent garnet, assuming that the effective depth equals its total thickness of $7.5~\mu$m. Strong enhancement of the efficiency at the SPP resonance in the Au/YIG:Co sample is in a striking contrast with the flat spectral dependence on a bare garnet. It is seen that the SPP excitation results in the 6-fold enhancement of the specific amplitude of the magnetization precession at the resonance.

It is worth emphasizing the similarities and differences between the systems studied here and in our recent work \cite{Chekhov2018}. In both experiments, strong amplification of the spin dynamics is inherently related to the SPP-driven localization of light at the interface. Together with the enhancement of the SPP electric fields, this effect is generic for the entire class of metal-dielectric plasmonic heterostructures and can be further optimized for better performance. From the photonic point of view, another common important impact of the SPP excitation consists in the amplification of the out-of-plane projection of the electric field $E_z$ which is otherwise suppressed in the high-$n$ dielectric.

Yet, the tensorial character of Eq.~\ref{eq1} reveals important differences between the two cases. In general, the following form for the effective opto-magnetic field $H_{\rm eff}$ can be derived\cite{KimelZvezdin}:

\begin{equation}
    H_{{\rm eff},i}(0)\propto\alpha_{ijk} E_j(\omega)E_k^*(-\omega)+\chi_{ijkl} E_j(\omega)E_k^*(-\omega)M_l(0)+c.c.,
    \label{eq3}
\end{equation}
where higher-order (in $M$) terms are neglected, and $\alpha_{ijk}$ and $\chi_{ijkl}$ are the antisymmetric and symmetric susceptibility tensors, respectively\cite{LL5}. The symmetry determines their dependence on the phase shift $\varphi$ and thus on the polarization of light. The inverse Faraday effect captured by the first term in Eq.~\ref{eq3} requires $\varphi\neq0$ which can be realized employing either circularly polarized light or SPP excitation\cite{Khokhlov2012,Chekhov2018}. On the contrary, the photo-magnetic effect is present even in the absence of SPP, and moreover, the purely SPP-driven photo-magnetic contribution vanishes due to $\varphi_{\rm SPP}=\pi/2$ (cf. Eq.~\ref{eq2}). Thus, the photo-magnetic SPP-mediated mechanism largely consists in the strong localization and enhancement of the electric field at the interface, whereas the excitation magnitude $\propto\mathcal{E}$ is determined by the optical interference of the SPP and incident fields. Similarly, the reversal of the precession phase  when comparing magneto-plasmonic crystals with the bare garnet, as seen in Fig.~\ref{fgr:2}, can also be attributed to this interference of the electromagnetic fields.

The excitation mechanism of the spin dynamics here originates in the Co doping of YIG, enabling setting magnetization into motion through effective SPP-mediated photo-magnetic anisotropy field.
The photo-magnetic mechanism offers the largest angles of magnetization precession available up to date at non-destructive laser fluences. Transient $\theta_F$ values observed in our experiments correspond to about $5^{\circ}$ magnetization excursion from the equilibrium, 1-2~orders of magnitude larger than that found in Gd,Yb-doped BIG \cite{Chekhov2018} and LuIG\cite{Hansteen2005}. It can be conjectured that further amplification is feasible through structural optimization of the plasmonic geometry aimed at enhancing the excitation $\mathcal{E}$ and employing numerical simulations. 

From the perspective of magnetic data recording, the photo-magnetic switching through large-angle precession has been demonstrated exclusively in YIG:Co. The SPP photo-magnetic mechanism thus holds high potential for taking the all-optical magnetization switching onto the nanoscale. In our prototype system, the observed 6-fold amplification of the specific efficiency $\xi$ allows corresponding reducing of the laser fluence below the switching threshold. The extrapolation to nm-sized bits yields about only $\sim2$~aJ deposited energy per one bit, close to the fundamental thermodynamical limit for switching magnetic bits, thus corroborating the exceptional energy efficiency of the photo-magnetic nanosize switching.

\begin{acknowledgement}

The authors thank A. Kirilyuk (Nijmegen) and T. V. Murzina (Moscow) for their support.
This work has been funded by the National Science Centre Poland (Grant
No. DEC-2017/25/B/ST3/01305) and ERC Consolidator Grant TERAMAG No. 681917.

\end{acknowledgement}

\bibliography{refs}

\end{document}